\documentclass[11pt,a4paper,twocolumns]{revtex4}
\usepackage{amsmath}
\usepackage{graphicx}
\usepackage{fancyhdr}
\usepackage{calc}
\usepackage{amssymb}
\usepackage[sort&compress]{natbib}
\usepackage{setspace}
\usepackage{amsfonts}
\usepackage{commath}
\usepackage[innercaption]{sidecap}
\usepackage[framemethod=TikZ]{mdframed}
\usepackage{hyperref}
\usepackage{parskip}

\expandafter\def\expandafter\normalsize\expandafter{%
    \normalsize
    \setlength\abovedisplayskip{0pt}
    \setlength\belowdisplayskip{5pt}
    \setlength\abovedisplayshortskip{0pt}
    \setlength\belowdisplayshortskip{5pt}
}

\definecolor{Gray}{gray}{0.75}

\newmdenv[backgroundcolor=Gray, innerbottommargin = 3pt, innertopmargin = 3pt, leftmargin = 0pt, rightmargin = 0pt, linewidth = 0pt, roundcorner = 1 pt, innerleftmargin=3pt, innerrightmargin=3pt]{Frame}

\begin{document}

%%%%%%%%%%%%%%%%%%%%%%%%%%%%%%
%% MATHEMATICAL NEWCOMMANDS %%
%%%%%%%%%%%%%%%%%%%%%%%%%%%%%%

\definecolor{Blue}{RGB}{30,30,255}

\setlength\arraycolsep{12pt}
\renewcommand{\arraystretch}{1.5}

\newcommand{\av}[1]
{
\langle #1 \rangle
}

\newcommand{\R} %MM%
{
\mathbb R
}

\newcommand{\E} %SIS%
{
\mathbb E
}

\newcommand{\M} %ECO%
{
\mathbb M
}

\renewcommand{\P} %PD%
{
\mathbb P
}

\newcommand{\B} %MAK%
{
\mathbb B
}

\newcommand{\N} %NEURO%
{
\mathbb N
}

\newcommand{\m}[3]
{
#1_{#2 #3}
}

\newcommand{\D}[2]
{
\mathcal{D}_{#1 #2}
}

\renewcommand{\L}[2]
{
\mathcal{L}{(#2 \rightarrow #1)}
}

\newcommand{\T}[2]
{
T (#2 \rightarrow #1)
}

\title{Response times of nodes in a complex network environment - \\ two potential derivation tracks}

\author{Chittaranjan Hens$^{1}$, Uzi Harush$^{2}$, Simcha Haber$^{2}$, Reuven Cohen$^{2}$ \& Baruch Barzel$^{2,3,*}$}
\affiliation{
\begin{enumerate}
\item
Physics and Applied Mathematics Unit, Indian Statistical Institute, Kolkata, India
\item
Department of Mathematics, Bar-Ilan University, Ramat-Gan, Israel 
\item
Gonda Multidisciplinary Brain Research Center, Bar-Ilan University, Ramat-Gan, Israel
\end{enumerate}
\begin{itemize}
\item[\textbf{*}]
\textbf{Correspondence}: baruchbarzel@gmail.com
\end{itemize} 
}

\setstretch{1.2}
   
\begin{abstract}
\textbf{The spread of perturbative signals in complex networks is governed by the combined effect of the network topology and its intrinsic nonlinear dynamics. Recently, the resulting spreading patterns have been analyzed and predicted, shown to depend on a single scaling relationship, linking a node's weighted degree $S_i$ to its intrinsic response time $\tau_i$. The relevant scaling exponent $\theta$ can be analytically traced to the system's nonlinear dynamics. Here we show that $\theta$ can be obtained via two different derivation tracks, leading to seemingly different functions. Analyzing the resulting predictions, we find that, despite their distinct form, they are fully consistent, predicting the exact same scaling relationship under potentially diverse types of dynamics.}
\end{abstract}
   
\maketitle

In Ref.\ \cite{Hens2019} we seek the patterns of signal propagation in a complex network environment. We begin with an $N$ node dynamic system, whose activities $x_i(t)$ ($i = 1,\dots,N$) are driven by \cite{Barzel2013,Barzel2015,Harush2017,Hens2019}

\begin{equation}
\dod{x_i}{t} = M_0(x_i) + \sum_{j = 1}^N A_{ij} M_1\big( x_i(t) \big) M_2\big( x_j(t) \big),
\label{Dynamics}
\end{equation}

where $\m Aij$ represents the $N \times N$ weighted adjacency matrix, and the nonlinear functions $M_0(x), M_1(x)$ and $M_2(x)$ capture the system's internal nonlinear mechanisms. To track the spread of perturbative signals we seek the response time $\tau_i$ of all nodes $i = 1,\dots,N$ to a directly incoming signal. We find that these response times are characterized by a single universal exponent $\theta$ via

\begin{equation}
\tau_i \sim S_i^{\theta}
\label{TauSi}
\end{equation}

where $S_i = \sum_{j = 1}^N \m Aij$ is node $i$'s weighted degree. Hence $\theta$ links a well-mapped topological feature, degree $S_i$, to its consequent dynamic observable, response time $\tau_i$. Averaging over all nodes with degree around $S$, \textit{i.e}.\ $S_i \in (S, S + \dif S)$, we rewrite (\ref{TauSi}) as

\begin{equation}
\tau(S) \sim S^{\theta},
\label{TauS}
\end{equation} 

predicting the average response times of nodes in function of their weighted degree. 

To predict $\theta$ we link it to the dynamics (\ref{Dynamics}) via \cite{Hens2019}

\begin{equation}
Y \big( R^{-1}(x) \big) = \sum_{n = 0}^{\infty} C_n x^{\Gamma(n)},
\label{YRx}
\end{equation}

where 

\begin{equation}
Y(x) = \left( \dod{ \big[ M_1(x) R(x) \big]}{x} \right)^{-1},
\label{Yx}
\end{equation}

$R(x) = -M_1(x)/M_0(x)$ and $R^{-1}(x)$ is its inverse function. The function (\ref{YRx}) is fully determined by the system's dynamics through $M_0(x)$ and $M_1(x)$. Its leading power $\Gamma(0)$ determines the exponent $\theta$ via

\begin{equation}
\theta = -2 - \Gamma(0).
\label{Theta}
\end{equation}

The detailed derivation of this result appears in Ref.\ \cite{Hens2019}.

Recently, an alternative derivation has been proposed, which leads to a seemingly distinct scaling. In this derivation track Eq.\ (\ref{Yx}) is substituted by 

\begin{equation}
Y(x) = \left( M_1(x) \dod{R(x)}{x} \right)^{-1},
\label{YxAlt}
\end{equation}

in which the function $M_1(x)$ is excluded from the $x$-derivative. Consequently, the power series expansion in (\ref{YRx}) is potentially different, and hence $\theta$ in (\ref{Theta}) may deviate from the original prediction of Ref.\ \cite{Hens2019}.

\vspace{2mm}
\begin{Frame}
Here we show that the two derivation tracks are, in fact, one and the same, exhibiting identical leading power $\Gamma(0)$, and therefore, predicting the exact same scaling exponent $\theta$. This self consistency of the two alternative derivations, provides further verification for the validity and universal nature of Eq.\ (\ref{TauS}). It also presents a solution to determine between the two alternative derivations, stating that non is preferable over the other - as they lead to indistinguishable results.
\end{Frame}

\section{Comparing the two tracks}

To systematically compare the two derivation tracks we refer to their point of bifurcation. Hence, we follow the original derivation, leading to (\ref{Yx}), where it is shown that \cite{Hens2019}  

\begin{equation}
\dfrac{1}{\tau(S)} = 
\dfrac{1}{\lambda^2} 
\left[
c_1 R \big( R^{-1}(\lambda) \big) M_1^{\prime} \big( R^{-1}(\lambda) \big) +
c_2 M_1 \big( R^{-1}(\lambda) \big) R^{\prime} \big( R^{-1}(\lambda) \big)
\right],
\label{TwoTerms}
\end{equation}

in which $\lambda \sim 1/S$ approaches zero in the limit of large $S$, and $c_1, c_2$ are arbitrary coefficients (see Supp.\ Eq.\ 1.34 in Red. \cite{Hens2019}). In \textbf{Track I} the next step is to arbitrarily set $c_1 = c_2 = 1$. This is motivated by the fact that these coefficients do not contribute to the scaling, and hence their specific value is of no importance. Once set to unity we can transform the r.h.s.\ of (\ref{TwoTerms}) into the product derivative of Eq.\ (\ref{Yx}), obtaining

\begin{equation}
\dfrac{1}{\tau(S)} = 
\dfrac{1}{\lambda^2} \dfrac{1}{Y \big( R^{-1}(x) \big)},
\label{Yone}
\end{equation}

and, consequently, the scaling in (\ref{Theta}). 

The alternative \textbf{Track II} sets $c_1 = 0$, preserving only the second term of (\ref{TwoTerms}), arriving at the single derivative of Eq.\ (\ref{YxAlt}), in which only $R(x)$ is differentiated. This track, therefore, predicts $\tau(S)$ via (\ref{Yone}), but with $Y(x)$ taken from (\ref{YxAlt}) instead of from (\ref{Yx}). To analyze the relationship between the two derivation tracks, we rewrite Eq.\ (\ref{TwoTerms}) as

\begin{equation}
\dfrac{1}{\tau(S)} = 
\dfrac{1}{\lambda^2} 
\Big(
c_1 Z_1(\lambda)  + c_2 Z_2(\lambda)
\Big),
\label{Z1Z2}
\end{equation}  

where

\begin{eqnarray}
Z_1(\lambda) &=& R \big( R^{-1}(\lambda) \big) M_1^{\prime} \big( R^{-1}(\lambda) \big)
\label{Z1}
\\[3pt]
Z_2(\lambda) &=& M_1 \big( R^{-1}(\lambda) \big) R^{\prime} \big( R^{-1}(\lambda) \big)
\label{Z2}
\end{eqnarray}

represent the two terms comprising the product derivative of (\ref{TwoTerms}). Using this description, we can summarize the difference between the two proposed derivations as

\begin{itemize}
\item
\textbf{Track I}.\ As appears in the original derivation \cite{Hens2019}, we take $c_1 \ne 0$, and hence keep both terms of Eq.\ (\ref{Z1Z2}).
\item
\textbf{Track II}.\ Setting the coefficient $c_1$ in Eq.\ (\ref{Z1Z2}) to zero, remaining only with the second term $Z_2(\lambda)$.
\end{itemize}

Mathematically speaking, \textbf{Track II} is only justifiable if $c_1$ is truly \textit{equal} to zero, while \textbf{Track I} is insensitive to the specific value of $c_1$, relevant even if, \textit{e.g.}, $c_1 \approx 0$. It is for this reason that we find \textbf{Track I} more general and rigorous. However, the focus here is on the \textit{outcome} of the two tracks, not their rigorousness.  

\textbf{Is there a difference between the Tracks?} Clearly Eq.\ (\ref{Z1Z2}) provides a different \textit{value} for $\tau(S)$, depending on $c_1$. Yet, our focus here is not on the specific \textit{value}, but rather on the \textit{asymptotic scaling} $\theta$. This is determined only by the power-series expansion of $Z_1(\lambda)$ vs.\ that of $Z_2(\lambda)$ in the limit $\lambda \rightarrow 0$, \textit{i.e}.\ $S \rightarrow \infty$. Writing

\begin{eqnarray}
\lim_{\lambda \rightarrow 0} Z_1(\lambda) &\sim& \lambda^{\alpha}
\label{Alpha}
\\[3pt]
\lim_{\lambda \rightarrow 0} Z_2(\lambda) &\sim& \lambda^{\beta},
\label{Beta}
\end{eqnarray}

we distinguish between three different scenarios:   

\begin{itemize}
\item[]
\textbf{\color{blue} Case 1}.\ $\alpha = \beta$.\ In this case both terms in (\ref{Z1Z2}) provide the same scaling, and hence \textbf{Track I} agrees with \textbf{Track II}. 
\item[]
\textbf{\color{blue} Case 2}.\ $\alpha > \beta$.\ Here in the limit $\lambda \rightarrow 0$ we have $Z_1(\lambda) \ll Z_2(\lambda)$, Eq.\ (\ref{Z1Z2}) is dominated by $Z_2(\lambda)$ independent of $c_1$, and therefore, once again, we have \textbf{Track I} and \textbf{Track II} in full agreement.
\item[]
\textbf{\color{blue} Case 3}.\ $\alpha < \beta$.\ This is the only scenario where $Z_1(\lambda)$ dominates Eq.\ (\ref{Z1Z2}). Here setting $c_1 = 0$ will leave only the $Z_2(\lambda)$ term and its associated exponent $\beta$, while $c_1 \ne 0$ will predict the dominance of $\alpha$, proposing a potential discrepancy between \textbf{Tracks I} and \textbf{II}.
\end{itemize}

Fortunately, we find below that \textbf{Case 3} is prohibited, and therefore \textbf{Track I}, offered in Ref.\ \cite{Hens2019}, provides the same prediction for $\theta$ as the suggested alternative \textbf{Track II}. 

\section{The scaling of $Z_1(\lambda)$ and $Z_2(\lambda)$}

To calculate $\alpha$ and $\beta$ in (\ref{Alpha}) and (\ref{Beta}) let us first assume that $M_1(x)$ is \textit{not} constant. Indeed, in case $M_1(x) = \rm{Const}$ the two tracks trivially coincide as (\ref{Yx}) and (\ref{YxAlt}) are identical. Hence we can write (to leading power)

\begin{equation}
M_1(x) \sim x^{\Phi},
\label{M1x}
\end{equation}

with $\Phi \ne 0$. The specific case where the leading power is $\Phi = 0$, and hence the \textit{next} leading power is required, is treated separately below. Following Eq.\ (\ref{M1x}) we have the leading power of $M^{\prime}(x)$ as

\begin{equation}
M_1^{\prime}(x) \sim x^{\Phi - 1}.
\label{M1Primex}
\end{equation}

Next we extract the power-series (Hahn expansion \cite{Hahn1995}) of $R^{-1}(x)$ around $x = 0$ as 

\begin{equation}
R^{-1}(x) = \sum_{n = 0}^{\infty} C_n x^{\Omega(n)},
\label{Rx}
\end{equation}

in which $\Omega(n)$ is a sequence of real powers, increasing with $n$. For small $x$, $R^{-1}(x)$ is dominated by the leading, \textit{i.e.\ smallest}, powers $\Omega(0), \Omega(1), \dots$. To obtain the powers $\alpha$ and $\beta$ in (\ref{Alpha}) and (\ref{Beta}), we must express the near-zero behavior of each of the terms in (\ref{Z1}) and (\ref{Z2}): $M_1(R^{-1}(x)), M_1^{\prime}(R^{-1}(x)), R(R^{-1}(x))$ and $R^{\prime}(R^{-1}(x))$. This depends on the exponents $\Phi$ in (\ref{M1x}) and $\Omega(n)$ in (\ref{Rx}). As we show below, we must first consider the value of $\Omega(0)$, specifically, whether it is positive, negative or zero (Fig.\ 1).

\begin{figure}
\begin{center}
\includegraphics[width = 16cm]{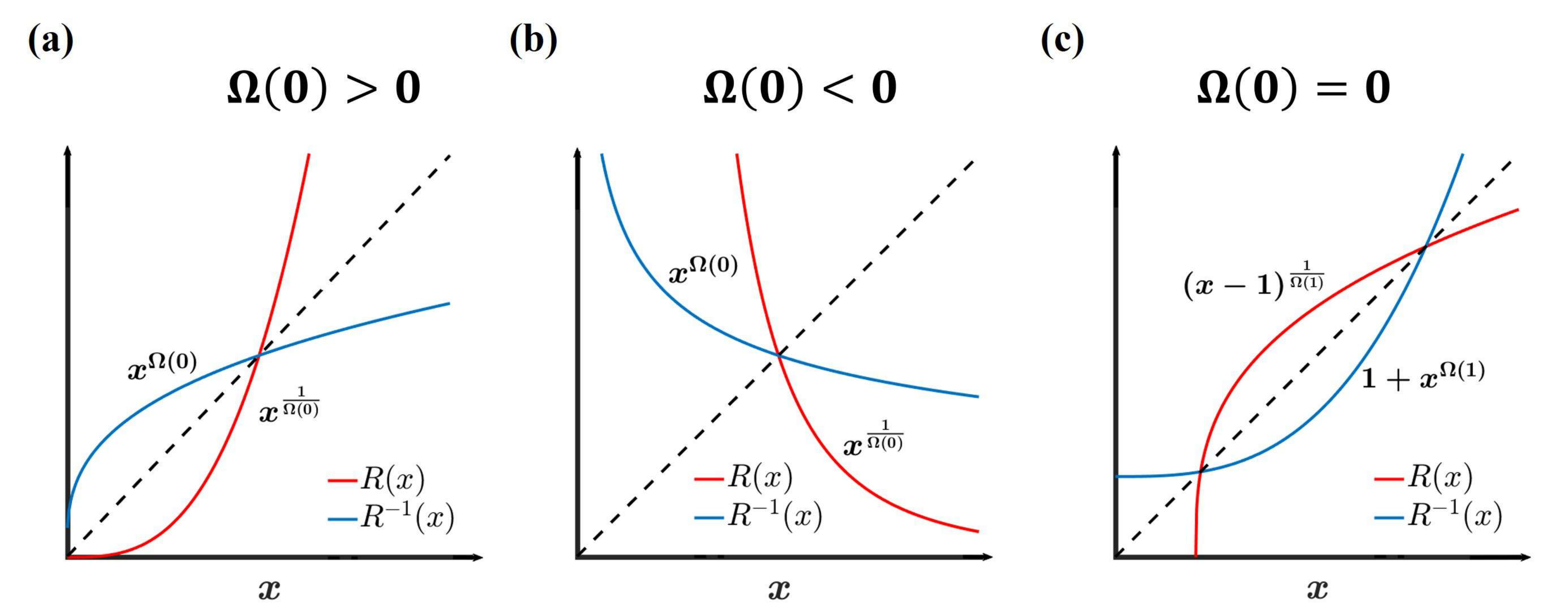}
\caption{{\color{blue} \textbf{The different behaviors of $R^{-1}(x)$}}.
The scaling of $Z_1$ and $Z_2$ in (\ref{Z1}) and (\ref{Z2}) depends on the limit $\lim_{x \rightarrow 0} R^{-1}(x)$. Using power series analysis we write $R^{-1}(x) = C_0 x^{\Omega(0)} + C_1 x^{\Omega(1)} + \dots$, and take the limit $x \rightarrow 0$, such that only the leading powers are significant. We distinguish between three different scenarios: 
{\color{blue} \textbf{(a)} $\Omega(0) > 0$}.\ Here we have $R^{-1}(x \rightarrow 0) \sim x^{\Omega(0)} \rightarrow 0$ (blue). This, by inversion, provides $R(x) \sim x^{1/\Omega(0)}$ (red), both limits relevant in the $x \rightarrow 0$ regime.\
{\color{blue} \textbf{(b)} $\Omega(0) < 0$}.\ Under a negative leading power we have $R^{-1}(x \rightarrow 0) \sim x^{\Omega(0)} \rightarrow \infty$. This provides $R(x) \sim x^{1/\Omega(0)}$, this time in the limit $x \rightarrow \infty$, instead of $x \rightarrow 0$.\
{\color{blue} \textbf{(c)} $\Omega(0) = 0$}.\ Here $R^{-1}(x \rightarrow 0)$ approaches a constant $C_0$ - hence, inversion helps characterize the original function $R(x)$ around $x \rightarrow C_0$, as $R(x) \sim [(x - C_0)/C_1]^{1/\Omega(1)}$, a shifted polynomial. In display we set both coefficients $C_0, C_1$ to unity, for simplicity. In all panels the dashed line represents $y = x$. 
}
\end{center}
\end{figure}

{\color{blue} \textbf{1.\ The case where $\Omega(0) > 0$} (Fig.\ 1a)}.\ If the leading power in (\ref{Rx}) is positive, we have in the limit $x \rightarrow 0$, $R^{-1}(x) \sim x^{\Omega(0)}$ (blue), providing $R(x) \sim x^{1/\Omega(0)}$ (red), and consequently $R^{\prime}(x) \sim x^{-1 + 1/\Omega(0)}$. We can therefore calculate each of the terms comprising $Z_1(\lambda)$ and $Z_2(\lambda)$ in (\ref{Z1}) and (\ref{Z2}) as

\begin{eqnarray}
M_1 \big( R^{-1} (\lambda) \big) &\sim& \Big( \lambda^{\Omega(0)} \Big)^{\Phi} = \lambda^{\Phi \Omega(0)}
\label{Terms11}
\\[3pt]
M_1^{\prime} \big( R^{-1} (\lambda) \big) &\sim& \Big( \lambda^{\Omega(0)} \Big)^{\Phi - 1} = \lambda^{\Phi \Omega(0) - \Omega(0)}
\label{Terms12}
\\[3pt]
R^{\prime} \big( R^{-1} (\lambda) \big) &\sim& \Big( \lambda^{\Omega(0)} \Big)^{-1 + 1/\Omega(0)} = \lambda^{1 - \Omega(0)}.
\label{Terms13}
\end{eqnarray}  

Collecting all the terms, and using the fact that $R(R^{-1}(\lambda)) = \lambda$ we obtain

\begin{eqnarray}
Z_1(\lambda) &\sim& \lambda^{1 + \Phi \Omega(0) - \Omega(0)}
\label{Z1Case1} 
\\[3pt]
Z_2(\lambda) &\sim& \lambda^{1 + \Phi \Omega(0) - \Omega(0)},
\label{Z2Case1}
\end{eqnarray}

and hence we observe $\alpha = \beta$ in (\ref{Alpha}) and (\ref{Beta}). This corresponds to \textbf{Case 1} above, in which it is guaranteed that \textbf{Tracks I} and \textbf{II} yield identical $\theta$.

\textbf{Example}.\ As an example we consider the dynamic model

\begin{equation}
\dod{x_i}{t} = -Bx_i + \sum_{j = 1}^N \m Aij x_i^{\frac{3}{2}} f(x_j),
\label{Model1}
\end{equation}

for which we have $M_1(x) \sim x^{\frac{3}{2}}$, \textit{i.e}.\ $\Phi = 3/2$ in (\ref{M1x}). For this model we have $R(x) \sim x^{1/2}$, providing $\Omega(0) = 1/2 > 0$, and correspondingly $R^{\prime}(x) \sim x^{-1/2}$ and $R^{-1}(x) \sim x^2$. Using \textbf{Track I}'s Eq.\ (\ref{Yx}) we write 

\begin{equation}
Y\Big( R^{-1}(\lambda) \Big) = \left. \left( \dod{M_1 R}{x} \right)^{-1} \right|_{x = R^{-1}(\lambda)} = 
\left. \dfrac{1}{2x} \right|_{x = \lambda^2} \sim \lambda^{-2}, 
\label{Y1}
\end{equation}

providing $\Gamma(0) = -2$, and hence, through Eq.\ (\ref{Theta}), predicting  $\theta_{\rm{I}} = -2 - \Gamma(0) = 0$.

On the other hand, following \textbf{Track II } we use Eq.\ (\ref{YxAlt}) to write 

\begin{equation}
Y\Big( R^{-1}(\lambda) \Big) =  
\left. \left( M_1(x) \dod{R}{x} \right)^{-1} \right|_{x = R^{-1}(\lambda)} = 
\Big( M_1 \big( \lambda^2 \big) R^{\prime} \big( \lambda^2 \big) \Big)^{-1} = \lambda^{-2}, 
\label{Tau1I}
\end{equation}

the same leading power, predicting again that $\theta_{\rm{II}} = 0$. Therefore, as predicted, we observe that 

\begin{equation}
\theta_{\rm{I}} = \theta_{\rm{II}}
\end{equation}

namely that both \textbf{Tracks I} and \textbf{II} are in perfect agreement.

\vspace{2mm}
{\color{blue} \textbf{2.\ The case where $\Omega(0) < 0$} (Fig.\ 1b)}.\
When the leading power in (\ref{Rx}) is negative, we have $R^{-1}(x) \sim x^{\Omega(0)} \rightarrow \infty$ for $x \rightarrow 0$ (blue). This corresponds to $R(x) \sim x^{1/\Omega(0)}$ in the limit $x \rightarrow \infty$ (red). Hence, in this case, the inverse $R^{-1}(x)$ helps characterize the asymptotic behavior of $R(x)$ in the $x \rightarrow \infty$ limit, as opposed to the $x \rightarrow 0$ limit, obtained in the $\Omega(0)$ positive scenario. We therefore obtain the large $x$ limit of the derivative $R^{\prime}(x)$ as $R^{\prime}(x) \sim x^{1/\Omega(0) - 1}$, helping us, once again, construct all the terms of (\ref{Z1}) and (\ref{Z2}) as

\begin{eqnarray}
M_1 \big( R^{-1} (\lambda) \big) &\sim& \Big( \lambda^{\Omega(0)} \Big)^{\Phi} = \lambda^{\Phi \Omega(0)}
\label{Terms21}
\\[3pt]
M_1^{\prime} \big( R^{-1} (\lambda) \big) &\sim& \Big( \lambda^{\Omega(0)} \Big)^{\Phi - 1} = \lambda^{\Phi \Omega(0) - \Omega(0)}
\label{Terms22}
\\[3pt]
R^{\prime} \big( R^{-1} (\lambda) \big) &\sim& \Big( \lambda^{\Omega(0)} \Big)^{-1 + 1/\Omega(0)} = \lambda^{1 - \Omega(0)}.
\label{Terms23}
\end{eqnarray}  

The result is identical to (\ref{Terms11}) - (\ref{Terms13}), with the only difference that now $\Omega(0)$ is negative, and the argument $R^{-1}(\lambda)$ appearing in all the three functions approaches infinity and not zero. Hence (\ref{Terms21}) - (\ref{Terms23}) examine the behavior of $M(x)$ and $R(x)$ at the $x \rightarrow \infty$ limit. This is consistent with the fact that under $\Omega(0) < 0$ the steady state activities $x(S)$ scale positively with $S$ \cite{Barzel2013,Barzel2015}, and therefore the limit $S \rightarrow \infty$, relevant for our asymptotic scaling analysis, is associate with the \textit{large} $x$ nodes. In any case, what matters in the present context is that once collecting all the terms, we, again, arrive at 

\begin{equation}
\alpha = \beta = 1 + \Phi \Omega(0) - \Omega(0),
\label{AlphaBeta}
\end{equation}

satisfying \textbf{Case 1}, in which \textbf{Tracks I} and \textbf{II} remain consistent. 

\textbf{Example}.\ Here we consider

\begin{equation}
\dod{x_i}{t} = -Bx_i^2 + \sum_{j = 1}^N \m Aij x_i^{\frac{3}{2}} f(x_j),
\label{Model2}
\end{equation}

for which we have $\Phi = 3/2$ in (\ref{M1x}), and $R(x) \sim x^{-1/2}$. We now have $R^{\prime}(x) \sim x^{-3/2}$ and $R^{-1}(x) \sim x^{-2}$, \textit{i.e}.\ $\Omega(0) = -2 < 0$. Via \textbf{Track I} we write 

\begin{equation}
Y\Big( R^{-1}(\lambda) \Big) = \left. \left( \dod{M_1 R}{x} \right)^{-1} \right|_{x = R^{-1}(\lambda)} = 
1 \sim \lambda^0, 
\label{Y2}
\end{equation}

obtaining $\Gamma(0) = 0$, which using (\ref{Theta}) provides $\theta_{\rm{I}} = -2 - \Gamma(0) = -2$. Under \textbf{Track II} we have
 
\begin{equation}
Y\Big( R^{-1}(\lambda) \Big) =  
\left. \left( M_1(x) \dod{R}{x} \right)^{-1} \right|_{x = R^{-1}(\lambda)} = 
M_1 \big( \lambda^{-2} \big) R^{\prime} \big( \lambda^{-2} \big) = \lambda^0,
\label{Tau2I}
\end{equation}

again providing $\theta_{\rm{II}} = -2 - \Gamma(0) = -2$. Hence, once again, we observe that 

\begin{equation}
\theta_{\rm{I}} = \theta_{\rm{II}},
\end{equation}

with both \textbf{Tracks}, \textbf{I} and \textbf{II}, in full agreement.

\vspace{2mm}
{\color{blue} \textbf{3.\ The case where $\Omega(0) = 0$} (Fig.\ 1c)}.\
The third case we consider is when the leading power in in the expansion of $R^{-1}(x)$ vanishes. This represents the scenario where Eq.\ (\ref{Rx}) takes the form

\begin{equation}
R^{-1}(x) \sim C_0 + C_1 x^{\Omega(1)} + \dots \sim 1 + x^{\Omega(1)},
\label{Rx0}
\end{equation}

where in the last step we omitted the coefficients $C_0, C_1$, for simplicity. The second leading power is, by definition, greater than $\Omega(0)$, namely we have $\Omega(1) > 0$. Here, $R^{-1}(x)$ approaches a constant at $x \rightarrow 0$ (blue), expressed in the inverted $R(x)$ by a shifted polynomial of the form $R(x) \sim (x - 1)^{1/\Omega(1)}$ (red). Consequently $R^{\prime}(x) \sim (x - 1)^{1/\Omega(1) - 1}$. The terms to construct $Z_1(\lambda)$ and $Z_2(\lambda)$ become

\begin{eqnarray}
M_1 \big( R^{-1} (\lambda) \big) &\sim& \Big( 1 + \lambda^{\Omega(1)} \Big)^{\Phi} \sim 1 + \Phi \lambda^{\Omega(1)} + \dots \sim \lambda^0
\label{Terms31}
\\[3pt]
M_1^{\prime} \big( R^{-1} (\lambda) \big) &\sim& \Big( 1 + \lambda^{\Omega(1)} \Big)^{\Phi - 1} \sim 1 + (\Phi - 1) \lambda^{\Omega(1)} + \dots \sim \lambda^0
\label{Terms32}
\\[3pt]
R^{\prime} \big( R^{-1} (\lambda) \big) &\sim& \Big( 1 + \lambda^{\Omega(1)} - 1 \Big)^{1/\Omega(1) - 1} \sim \lambda^{1 - \Omega(1)},
\label{Terms33}
\end{eqnarray}  

and therefore

\begin{eqnarray}
Z_1(\lambda) &\sim& \lambda
\label{Z1Case3} 
\\[3pt]
Z_2(\lambda) &\sim& \lambda^{1 - \Omega(1)}.
\label{Z2Case3}
\end{eqnarray}

Using this in (\ref{Alpha}) and (\ref{Beta}) we obtain $\alpha = 1$ and $\beta = 1 - \Omega(1) < 1$ (since $\Omega(1) > 0$). Consequently, we have $\alpha > \beta$, satisfying \textbf{Case 2}, which once again predicts an agreement between \textbf{Track I} and \textbf{Track II}.

\textbf{Example}.\ 
To examine this class, the natural choice is to set $M_1(x) \sim \rm{Const}$, \textit{i.e}.\ $\Phi = 0$, however, in this case \textbf{Track I} trivially converges to \textbf{Track II}, therefore, we \textit{challenge} ourselves by constructing a non-trivial model

\begin{equation}
\dod{x_i}{t} = B - x_i^2 + \sum_{j = 1}^N \m Aij (1 - x_i) f(x_j),
\label{Model3}
\end{equation}

in which $M_0(x) = B - x^2$ and $M_1(x) = 1 - x$. Here 

\begin{eqnarray}
R(x) &=& - \dfrac{1 - x}{B - x^2}
\label{Model3R}
\\[3pt]
R^{\prime}(x) &=& \dfrac{x^2 - 2x + B}{x^4 - 2Bx^2 + B^2}  
\label{Model3Rprime}
\end{eqnarray}

and the inverse function is

\begin{equation}
R^{-1}(x) = \dfrac{1}{2x} 
\left(
-1 + \sqrt{1 + 4(Bx^2 + x)}
\right).
\label{Model3Rinverse}
\end{equation}

In the limit $\lambda \rightarrow 0$ we can approximate (\ref{Model3Rinverse}) by

\begin{equation}
R^{-1}(\lambda) \sim 1 + B \lambda + O(\lambda^2),
\label{Model3RinverseApprox}
\end{equation}

allowing us to evaluate the relevant approximations for $M_1(R^{-1}(\lambda))$ and $R^{\prime}(R^{-1}(\lambda))$, as required to calculate $\theta$ via both tracks. This time, we first begin with \textbf{Track II}, writing

\begin{equation}
Y\big( R^{-1}(\lambda) \big) =  
\left. \left( M_1(x) \dod{R}{x} \right)^{-1} \right|_{x = R^{-1}(\lambda)} = 
\left. \left(
(1 - x) 
\left( 
\dfrac{x^2 - 2x + B}{x^4 - 2Bx^2 + B^2}
\right) \right)^{-1} \right|_{x = 1 + B\lambda},
\label{Tau3I}
\end{equation}

where we have taken (\ref{Model3Rprime}) and (\ref{Model3RinverseApprox}) to express $R^{\prime}(x)$ and $R^{-1}(\lambda)$. Taking the limit where $\lambda \rightarrow 0$, the rational function on the r.h.s.\ of (\ref{Tau3I}) approaches a constant, leaving us with $Y(R^{-1}(\lambda)) \sim (B\lambda)^{-1}$, namely $\Gamma(0) = -1$, and hence 

\begin{equation}
\theta_{\rm II} = -2 - \Gamma(0) = -1.
\label{Tau3IFinal}
\end{equation}

Next we use \textbf{Track I}, writing

\begin{equation}
Y(x) = \left( \dod{M_1R}{x} \right)^{-1} = 
\left[
\dod{}{x} \left( (1 - x) \left( - \dfrac{1 - x}{B - x^2} \right) \right)  
\right]^{-1} = \dfrac{x^4 - 2Bx^2 + B^2}{2x^2 - 2(B + 1)x + 2B}.
\label{Y3}
\end{equation}

Taking (\ref{Model3RinverseApprox}) to approximate $R^{-1}(\lambda)$ we obtain

\begin{equation}
Y\Big( R^{-1}(\lambda) \Big) \approx \dfrac{1 - 2B + B^2}{(2B - 2B^2)\lambda} \sim \lambda^{-1},
\label{Y3Lambda}
\end{equation}

where we have only kept leading order terms in the limit $\lambda \rightarrow 0$. Equation (\ref{Y3Lambda}) provides us with $\Gamma(0) = -1$, leading to $\theta_{\rm{I}} = -2 - \Gamma(0) = -1$. Once again, even in these rather complex dynamics, we observe that 

\begin{equation}
\theta_{\rm{I}} = \theta_{\rm{II}},
\end{equation}

showing the complete agreement between the two derivation tracks.

\vspace{2mm}
{\color{blue} \textbf{4.\ The case where $\Phi = 0$}}.\
To complete our analysis there is one remaining scenario that has not yet been treated, captured by 

\begin{equation}
M_1(x) \sim C_0 + C_1 x^{\Phi(1)} + \dots \sim 1 + x^{\Phi(1)},
\label{M1x0}
\end{equation}

namely that the leading power in the expansion of $M_1(x)$ vanishes, and therefore its \textit{next} power $\Phi(1) > 0$ continues to play a role in determining $\theta$. The derivative is now

\begin{equation}
M_1^{\prime}(x) \sim x^{\Phi(1) - 1}.
\label{M1Primex0}
\end{equation}

Taking $R^{-1}(\lambda) \sim \lambda^{\Omega(0)}$, we recalculate the terms in (\ref{Terms11}) - (\ref{Terms13}) as

\begin{eqnarray}
M_1 \big( R^{-1} (\lambda) \big) &\sim& M_1 \Big( \lambda^{\Omega(0)} \Big) \sim 
1 + \lambda^{\Phi(1) \Omega(0)}
\label{Terms41}
\\[3pt]
M_1^{\prime} \big( R^{-1} (\lambda) \big) &\sim& \Big( \lambda^{\Omega(0)} \Big)^{\Phi(1) - 1} =
\lambda^{\Phi(1) \Omega(0) - \Omega(0)}
\label{Terms42}
\\[3pt]
R^{\prime} \big( R^{-1} (\lambda) \big) &\sim& \Big( \lambda^{\Omega(0)} \Big)^{-1 + 1/\Omega(0)} = 
\lambda^{1 - \Omega(0)},
\label{Terms43}
\end{eqnarray}  

valid under both negative or positive $\Omega(0)$. The case $\Omega(0) = 0$ remains the same as in (\ref{Terms31}) - (\ref{Terms33}), and therefore does not require specific treatment. Note that the only term that has changed due to $M_1(x)$'s vanishing leading power is the expression in Eq.\ (\ref{Terms41}), whose scaling is different compared with Eqs.\ (\ref{Terms11}) and (\ref{Terms21}) that were obtained under $M_1(x) \sim x^{\Phi}$. In the relevant limit, \textit{i.e}.\ $\lambda \rightarrow 0$, Eq.\ (\ref{Terms41}) is dominated either by the unity term ($\lambda^0$) or by the $\lambda^{\Phi(1) \Omega(0)}$ term, depending on which of the two powers is \textit{smaller}. This allows us to express Eq.\ (\ref{Terms41}) asymptotically as

\begin{equation}
M_1 \big( R^{-1} (\lambda) \big) \sim \lambda^{\min(0,\Phi(1) \Omega(0))},
\label{Terms41Min}
\end{equation}

keeping only the term raised to the minimal power, $0$ or $\Phi(1) \Omega(0)$. We can now collect all the terms, as we did in all previous cases, to construct $Z_1(\lambda)$ and $Z_2(\lambda)$ in (\ref{Z1}) and (\ref{Z2}), obtaining

\begin{eqnarray}
Z_1(\lambda) &\sim& \lambda^{1 + \Phi(1) \Omega(0) - \Omega(0)}
\label{Z1Case4} 
\\[3pt]
Z_2(\lambda) &\sim& \lambda^{1 + \min(0,\Phi(1) \Omega(0)) - \Omega(0)},
\label{Z2Case4}
\end{eqnarray}

for which the powers are

\begin{eqnarray}
\alpha &=& 1 + \Phi(1) \Omega(0) - \Omega(0)
\label{AlphaCase4} 
\\[3pt]
\beta &=& 1 + \min \big( 0,\Phi(1) \Omega(0) \big) - \Omega(0).
\label{BetaCase4}
\end{eqnarray}

Of course, it is guaranteed, by definition, that

\begin{equation}
\min \big( 0, \Phi(1) \Omega(0) \big) \le \Phi(1) \Omega(0)
\label{Minimum}
\end{equation}

and therefore that $\alpha \ge \beta$. This represents either \textbf{Case 1} or \textbf{Case 2}, but excludes \textbf{Case 3}, thus predicting, once again, the agreement of \textbf{Track I} and \textbf{Track II}.

\textbf{Example}. A classic example for a dynamics with $M_1(x)$ following the form of Eq.\ (\ref{M1x0}) is the SIS model \cite{Hufnagel2004}, for which $M_1(x) = 1 - x$. This dynamics has already been analyzed in Ref.\ \cite{Hens2019}, and one can readily confirm that both \textbf{Tracks} agree that it exhibits $\theta_{\rm I} = \theta_{\rm II} = -1$. Therefore, to expand the range of our validation we consider a \textit{new} type of dynamics of the form

\begin{equation}
\dod{x_i}{t} = -Bx_i + \sum_{j = 1}^N \m Aij \Big(1 - x_i^{\frac{1}{2}} \Big) f(x_j),
\label{Model4}
\end{equation}

capturing a potential generalization of the classic SIS model. Here we have $M_1(x) \sim 1 - x^{1/2}$, $R(x) = (1 - x^{1/2})/Bx$, and therefore

\begin{equation}
R^{-1}(x) = \dfrac{1}{2 B^2 x^2} 
\left(
1 + 2Bx - \sqrt{1 + 4Bx}
\right).
\label{Rinv4}
\end{equation}

Taking the relevant limit we obtain

\begin{equation}
\lim_{\lambda \rightarrow 0} R^{-1}(\lambda) = 1 - 2 B \lambda + O(\lambda^2),
\end{equation}

an expansion of the form (\ref{Rx}) with $\Omega(0) = 0$ and $\Omega(1) = 1$. Under \textbf{Track I} we have

\begin{equation}
Y(x) = \left( \dod{M_1 R}{x} \right)^{-1} = 
\left( \dod{}{x} \left( \dfrac{\Big(1 - x^{\frac{1}{2}} \Big)^2}{Bx} \right) \right)^{-1} = 
\dfrac{B x^2}{x^{\frac{1}{2}} - 1}.
\label{Y4}
\end{equation}

Using $R^{-1}(\lambda) \sim 1 - 2B\lambda$ we write

\begin{equation}
Y \big( R^{-1}(\lambda) \big) \sim 
\dfrac{B (1 - 2B\lambda)^2}{\sqrt{1 - 2B\lambda} - 1} \sim \lambda^{-1},
\label{YR4}
\end{equation}

providing $\Gamma(0) = -1$, and, as a result $\theta_{\rm I} = -2 - \Gamma(0) = -1$.

For \textbf{Track II} we write

\begin{eqnarray}
Y \big( R^{-1}(\lambda) \big) &=&  
\left. \left( M_1(x) \dod{R}{x} \right)^{-1} \right|_{x = R^{-1}(\lambda)} = 
\left. \left(
\Big(1 - x^{\frac{1}{2}} \Big) 
\left( 
\dfrac{x^{\frac{1}{2}} - 2}{2Bx^2} 
\right) \right)^{-1} \right|_{x = 1 - 2 B \lambda}
\nonumber \\[3pt]
&=&
\left[\left(
1 - \sqrt{1 - 2B\lambda}
\right)
\dfrac{\sqrt{1 - 2B\lambda} - 2}{2B(1 - 2B\lambda)^2} 
\right]^{-1},
\label{Tau4I}
\end{eqnarray}
 
which taking $\lambda \rightarrow 0$ provides

\begin{equation}
Y \big( R^{-1}(\lambda) \big) \sim \lambda^{-1},
\label{TauS4I}
\end{equation}

\textit{i.e}.\ $\theta_{\rm II} = -1$. Hence, also here $\theta_{\rm I} =\theta_{\rm II}$, reconfirming the consistency of \textbf{Track I} and \textbf{Track II}.

\vspace{2mm}
\begin{Frame}
Taken together, we arrive at an analytical proof, further strengthened by a broad array of dynamic examples, that the universal scaling exponent $\theta$ is independent of the derivation track, whether the original \textbf{Track I}, offered in Ref.\ \cite{Hens2019}, or the alternative \textbf{Track II}, suggested recently.
\end{Frame}

\clearpage
\renewcommand{\bibsection}{\section*{\color{Blue} REFERENCES}}

\bibliographystyle{unsrt}
%\bibliography{../../../../../../0Bibliography/bibliography}

%\begin{thebibliography}{10}

\end{document}